\documentstyle[12pt]{article}

\catcode`\@=11
\long\def\@makefntext#1{
\protect\noindent \hbox to 3.2pt {\hskip-.9pt  
$^{{\ninerm\@thefnmark}}$\hfil}#1\hfill}		

\def\@makefnmark{\hbox to 0pt{$^{\@thefnmark}$\hss}}  
	
\def\ps@myheadings{\let\@mkboth\@gobbletwo
\def\@oddhead{\hbox{}
\rightmark\hfil\ninerm\thepage}   
\def\@oddfoot{}\def\@evenhead{\ninerm\thepage\hfil
\leftmark\hbox{}}\def\@evenfoot{}
\def\sectionmark##1{}\def\subsectionmark##1{}}

\renewcommand{\thefootnote}{\fnsymbol{footnote}}

\newcounter{sectionc}\newcounter{subsectionc}\newcounter{subsubsectionc}
\renewcommand{\section}[1] {\vspace*{0.6cm}\addtocounter{sectionc}{1} 
\setcounter{subsectionc}{0}\setcounter{subsubsectionc}{0}\noindent 
	{\normalsize\bf\thesectionc. #1}\par\vspace*{0.4cm}}
\renewcommand{\subsection}[1] {\vspace*{0.6cm}\addtocounter{subsectionc}{1} 
	\setcounter{subsubsectionc}{0}\noindent 
	{\normalsize\it\thesectionc.\thesubsectionc. #1}\par\vspace*{0.4cm}}
\renewcommand{\subsubsection}[1]
{\vspace*{0.6cm}\addtocounter{subsubsectionc}{1}
	\noindent {\normalsize\rm\thesectionc.\thesubsectionc.\thesubsubsectionc. 
	#1}\par\vspace*{0.4cm}}

\newcounter{appendixc}
\newcounter{subappendixc}[appendixc]
\newcounter{subsubappendixc}[subappendixc]

\renewcommand{\appendix}[1] {\vspace*{0.6cm}
        \refstepcounter{appendixc}
        \setcounter{figure}{0}
        \setcounter{table}{0}
        \setcounter{equation}{0}
        \renewcommand{\thefigure}{\Alph{appendixc}.\arabic{figure}}
        \renewcommand{\thetable}{\Alph{appendixc}.\arabic{table}}
        \renewcommand{\theappendixc}{\Alph{appendixc}}
        \renewcommand{\theequation}{\Alph{appendixc}.\arabic{equation}}
        \noindent{\bf Appendix \theappendixc #1}\par\vspace*{0.4cm}}

\def\abstracts#1{{
	\centering{\begin{minipage}{12.2truecm}\footnotesize\baselineskip=12pt\noindent
	\centerline{\footnotesize ABSTRACT}\vspace*{0.3cm}
	\parindent=0pt #1
	\end{minipage}}\par}} 

\renewenvironment{thebibliography}[1]
	{\begin{list}{\arabic{enumi}.}
	{\usecounter{enumi}\setlength{\parsep}{0pt}
\setlength{\leftmargin 1.25cm}{\rightmargin 0pt}
	 \setlength{\itemsep}{0pt} \settowidth
	{\labelwidth}{#1.}\sloppy}}{\end{list}}

\newcounter{itemlistc}
\newcounter{romanlistc}
\newcounter{alphlistc}
\newcounter{arabiclistc}

\newcommand{\fcaption}[1]{
        \refstepcounter{figure}
        \setbox\@tempboxa = \hbox{\footnotesize Fig.~\thefigure. #1}
        \ifdim \wd\@tempboxa > 6in
           {\begin{center}
        \parbox{6in}{\footnotesize\baselineskip=12pt Fig.~\thefigure. #1}
            \end{center}}
        \else
             {\begin{center}
             {\footnotesize Fig.~\thefigure. #1}
              \end{center}}
        \fi}

\newcommand{\tcaption}[1]{
        \refstepcounter{table}
        \setbox\@tempboxa = \hbox{\footnotesize Table~\thetable. #1}
        \ifdim \wd\@tempboxa > 6in
           {\begin{center}
        \parbox{6in}{\footnotesize\baselineskip=12pt Table~\thetable. #1}
            \end{center}}
        \else
             {\begin{center}
             {\footnotesize Table~\thetable. #1}
              \end{center}}
        \fi}

\def\@citex[#1]#2{\if@filesw\immediate\write\@auxout
	{\string\citation{#2}}\fi
\def\@citea{}\@cite{\@for\@citeb:=#2\do
	{\@citea\def\@citea{,}\@ifundefined
	{b@\@citeb}{{\bf ?}\@warning
	{Citation `\@citeb' on page \thepage \space undefined}}
	{\csname b@\@citeb\endcsname}}}{#1}}

\newif\if@cghi
\def\cite{\@cghitrue\@ifnextchar [{\@tempswatrue
	\@citex}{\@tempswafalse\@citex[]}}
\def\citelow{\@cghifalse\@ifnextchar [{\@tempswatrue
	\@citex}{\@tempswafalse\@citex[]}}
\def\@cite#1#2{{$\null^{#1}$\if@tempswa\typeout
	{IJCGA warning: optional citation argument 
	ignored: `#2'} \fi}}

 1
 1
 1

\font\ninerm=cmr9

\begin{document}
\input psfig

\centerline{\normalsize\bf Martensitic Tweed and the Two--Way Shape--Memory 
Effect}

\centerline{\footnotesize James P.~Sethna}
\baselineskip=13pt
\centerline{\footnotesize\it Laboratory of Atomic and Solid State Physics,
Cornell University}
\baselineskip=12pt
\centerline{\footnotesize\it Ithaca, NY 14853-2501, USA}
\centerline{\footnotesize E-mail: sethna@lassp.cornell.edu}
\vspace*{0.3cm}
\centerline{\footnotesize and}
\vspace*{0.3cm}
\centerline{\footnotesize Christopher R.~Myers}
\baselineskip=13pt
\centerline{\footnotesize\it Cornell Theory Center, Cornell University,
Ithaca, NY 14853, USA}
\centerline{\footnotesize Originally presented at Los Alamos in April, 1995;
conference proceeding delayed.}

\vspace*{0.9cm}
\abstracts{We briefly introduce tweed, which is found above the
martensitic transition in a variety of shape-memory, high-T$_c$, and
other materials.  Based on our previous mapping of the problem onto
a spin-glass model, we conjecture that
the two-way shape-memory effect is due to tweed.}
 
\normalsize\baselineskip=15pt
\setcounter{footnote}{0}
\renewcommand{\thefootnote}{\alph{footnote}}
\section{What are Martensites?}

Martensites are crystalline materials which have undergone a crystalline
shape transition.\footnote{Not all changes in the crystal shape
give martensites: for example, if the volume changes too much,
the transition behaves quite differently and isn't called martensitic.}
For example, many metals are body-centered cubic at high temperature,
and change to face-centered cubic at low temperature: this can be done
through stretching the crystal along one of the three cubic axes
(figure 1).

We'll be using a simple, two-dimensional model for the martensitic
transition in this paper.  Our model will be a square lattice at
high temperatures, and a rectangular lattice at low temperatures
(with two {\it variants}: tall-and-thin and short-and-fat).

You can imagine that, when the material is only partly transformed,
there will be a lot of strain at the boundary!  (Half is stretched,
half not: how will it avoid tearing in between?)  It manages by
using the several different stretching directions!  Figure~2 shows
a photograph by Chunhwa Chu and Richard James (Dr. Chunhwa Chu, Ph.D Thesis,
University of Minnesota, 1993) of the domain patterns in their experiment.
The light region to the upper right might correspond to an undeformed,
high temperature cubic region; the striped region on the lower left
is composed of two variants, stretched in two different directions.
By making thin layers of two variants, the material manages to have no
net stretch at the boundary!  The different variants are
separated by what are called {\it twin} boundaries.

\begin{figure}
\vspace*{13pt}
\centerline{\psfig{figure=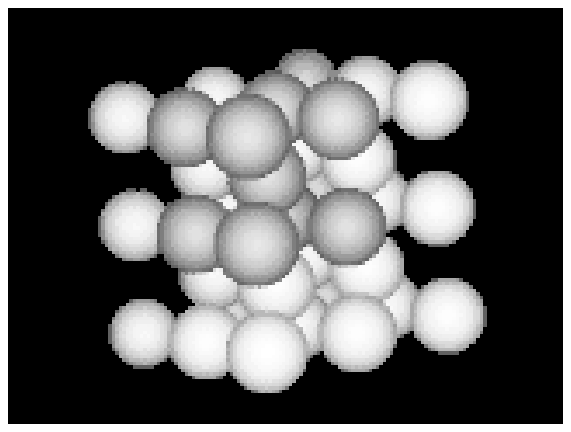,width=2.0truein}
~~\psfig{figure=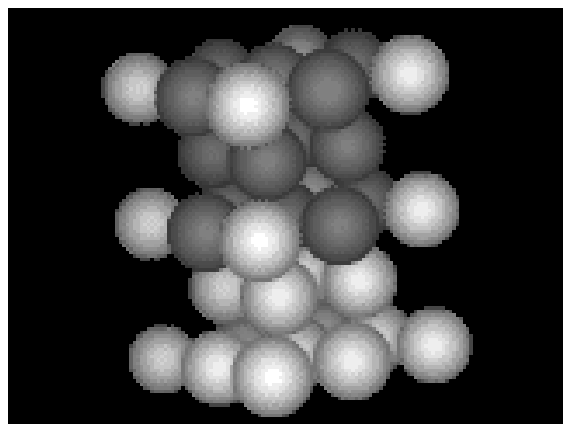,width=2.0truein}}
\fcaption{Body-Centered Cubic to Face-Centered Cubic.}
\label{fig:bcctofcc}
\end{figure}

\begin{figure}
\vspace*{13pt}
\centerline{\psfig{figure=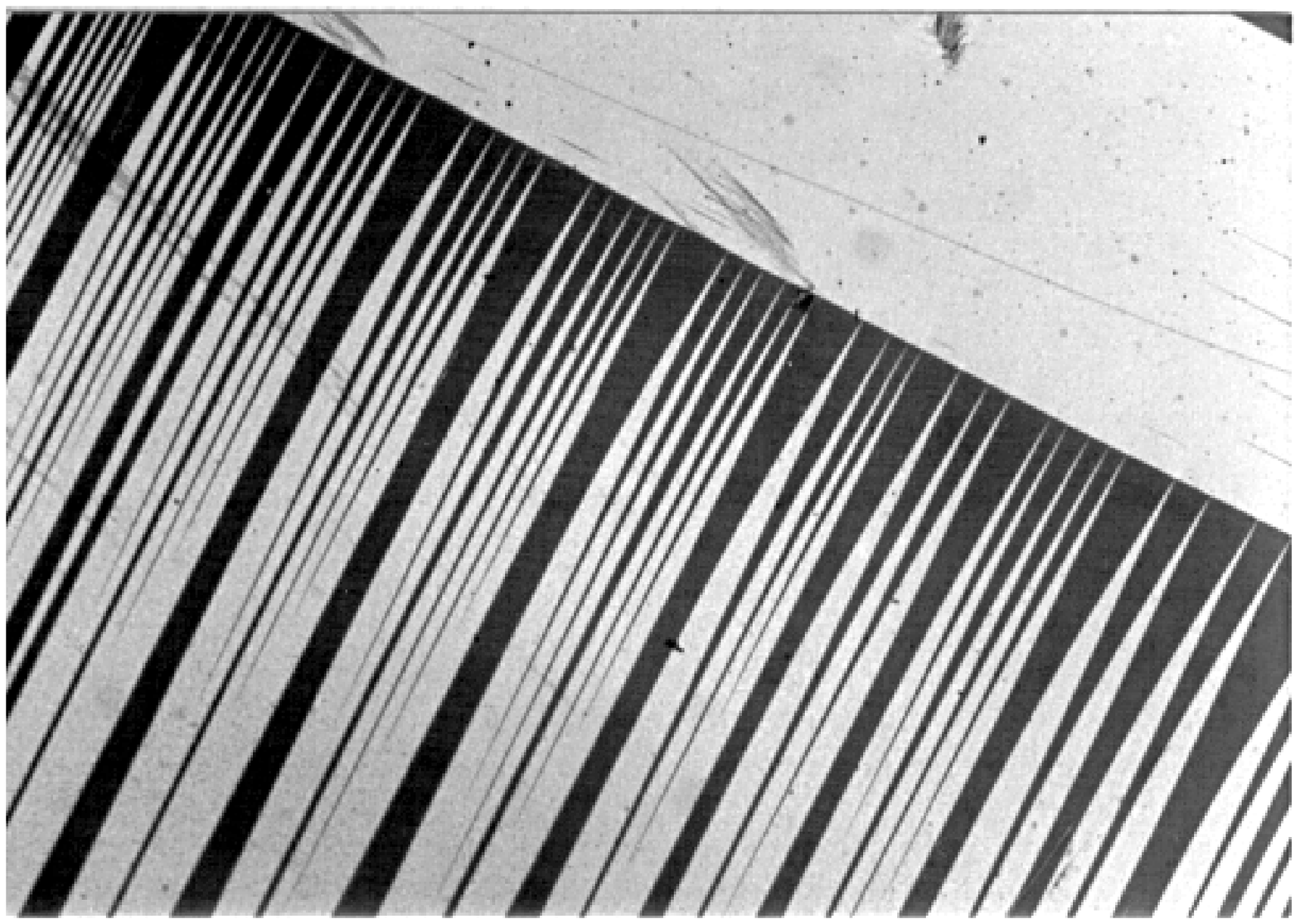,width=2.0truein}}
\fcaption{Martensitic Twins.  These are from Chu and James's
experiments in Cu-Al-Ni single crystals.}
\label{fig:chu}
\end{figure}

These twins are responsible for the {\it shape-memory effect}.
If one takes a teapot made of the cubic, high temperature
phase of a shape-memory martensite, and cools it through the
martensitic transition, it won't get tall and thin of course ---
it'll stay macroscopically the same shape, but now will be
made up of tiny slivers of various stretched variants, as in Figure~2.
The teapot will be much easier to dent in the twinned state,
since one no longer has to break bonds between atoms: one
can change the shape by moving the twin boundaries and changing
the relative amounts of differently stretched regions.
If one pulls our model horizontally, one gets more of the
tall-and-thin regions, and less short-and-fat (figure~3).

\begin{figure}
\vspace*{13pt}
\centerline{\psfig{figure=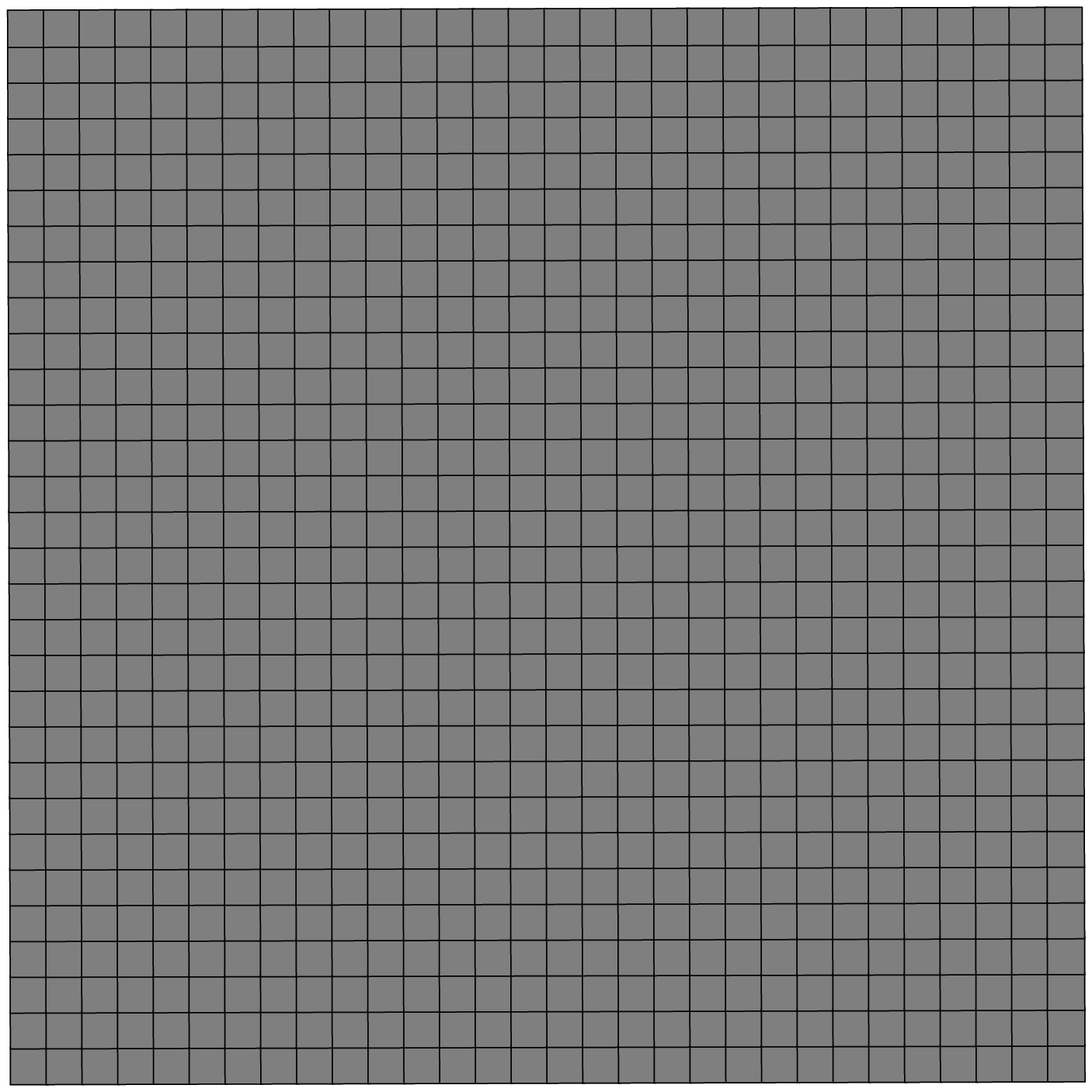,width=1.3truein}
~~\psfig{figure=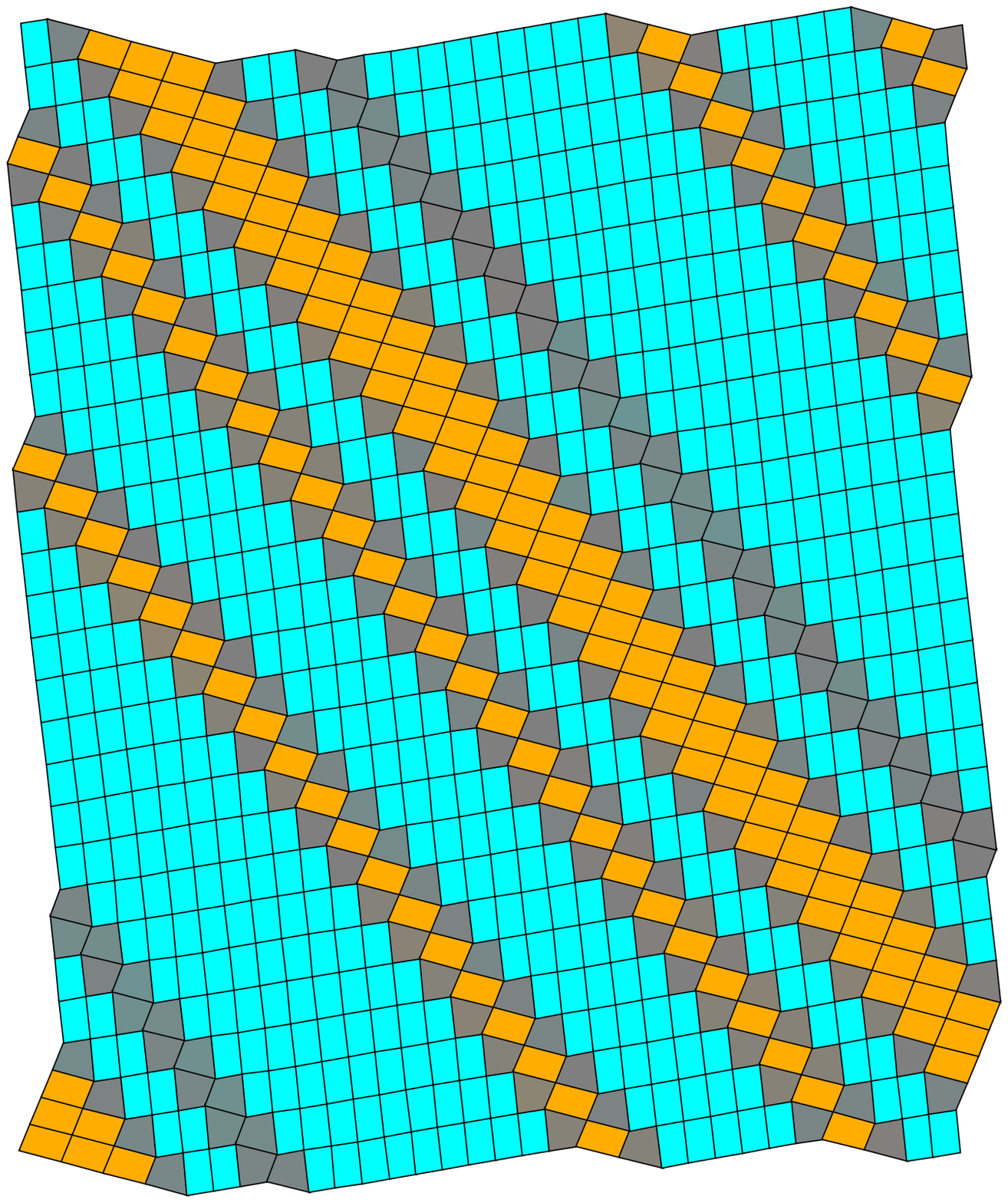,width=1.3truein}
~~\psfig{figure=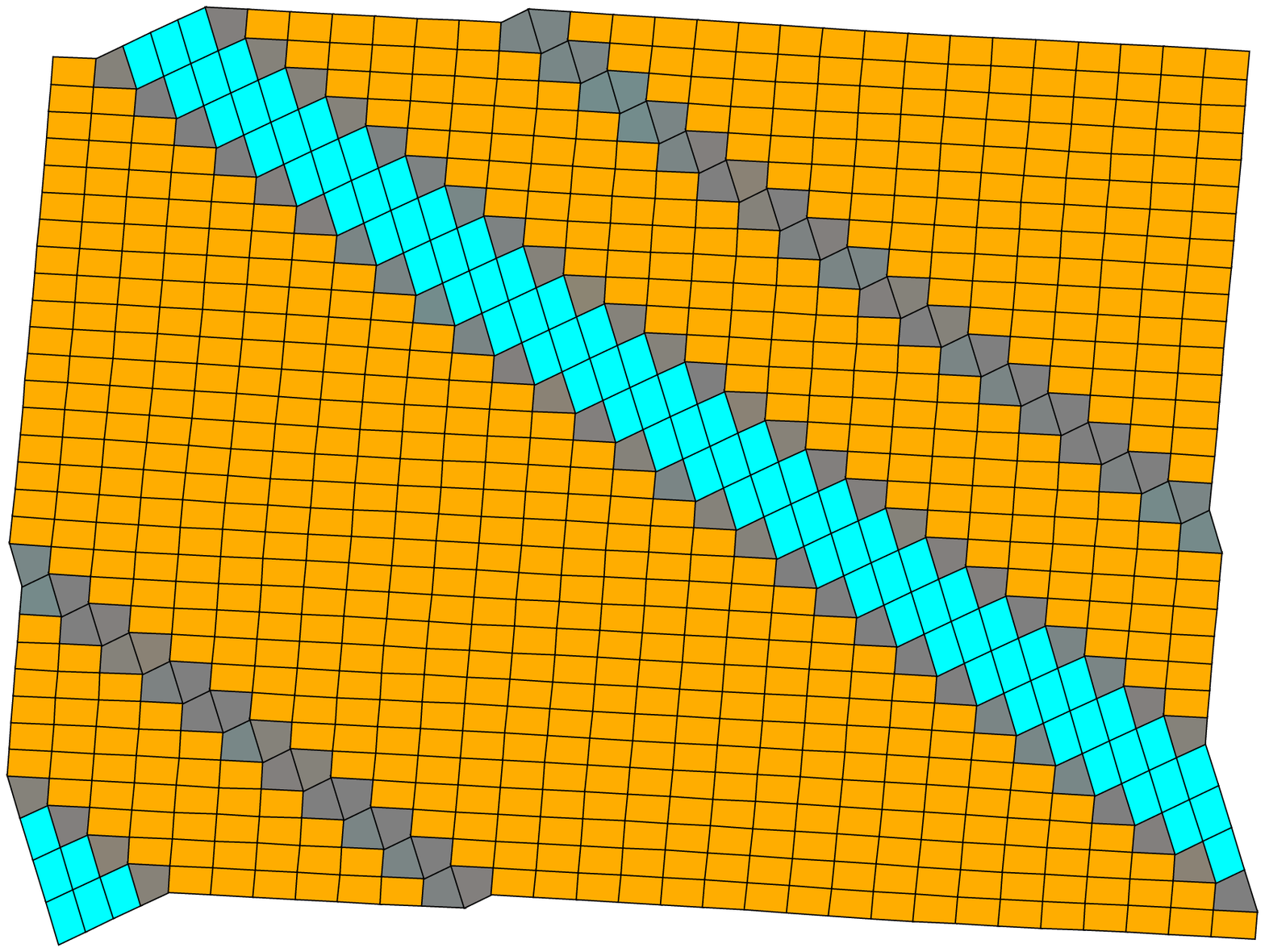,width=1.3truein}}
\fcaption{Teapot, Martensitic Teapot, and Strained
Martensitic Teapot.}
\label{fig:Teapot}
\end{figure}

Now, crush the teapot into a ball.  Heat the ball into
the phase where it wants to be cubic (square).  The ball
uncrumples back into a teapot!  Since no bonds were broken,
and since the twin boundaries which moved during the crumpling
vanish as the material is heated, it returns to its original
shape.  This is the (one-way) shape memory effect.

This paper is about the {\it two-way} shape memory effect.  If one
takes a shape-memory alloy, cools it, deforms it, and heats it,
it returns to its original shape (one-way memory).  If one
cycles through cooling/deforming/heating many times, each time
deforming in the same way, after tens or hundreds of cycles the
material deforms by itself when cooled.  This is quite a surprise!
How can the boring, cubic phase (grey on the left in figure~3)
remember how it should deform?

The community has proposals for what might store the memory.
Real martensites do break some bonds as they transform, leaving
dislocation lines: perhaps these remanents in the cubic
phase induce the two-way shape memory.  Also, these materials
often have inclusions or precipitates: perhaps these deform
during the cycling and induce future cycles to look like past ones.
We're proposing that these memories are stored in Tweed.

\section{What is Tweed?}

It turns out that, for tens to a hundred degrees above the martensitic
transition temperature, that these shape-memory alloys aren't boring,
cubic materials as suggested by Figure~3.  Instead, they exhibit
cross-hatched patterns reminiscent of threads in a tweed jacket!
Figure~4 shows a real picture of tweed: notice that the patchy regions
occur in stripes along the two diagonals.  Figure~5 shows our model\cite{Sivan}
of tweed, with realistic parameters (although the model, being two-dimensional,
isn't exactly completely realistic).

\begin{figure}
\vspace*{13pt}
\centerline{\psfig{figure=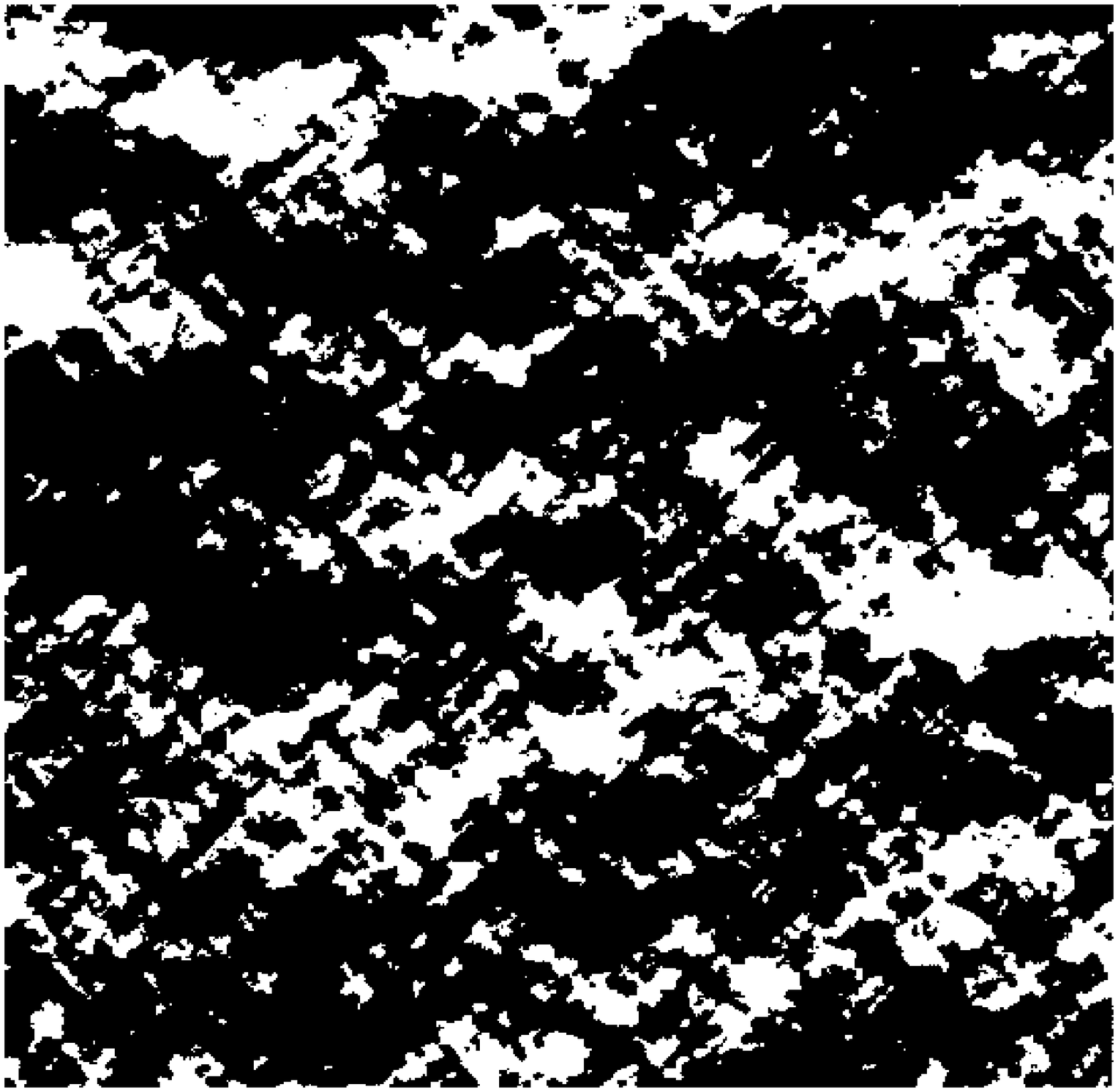,width=4.0truein}}
\bigskip
\fcaption{Tweed as experimentally observed in transmission electron
microscopy of a NiAl alloy (courtesy of Lee Tanner).  Tweed is identified by
its diagonal striations, which reflect some aperiodic lattice deformation
with correlations on the scale of some tens of atomic spacings. }
\label{fig:Tanner}
\end{figure}

\begin{figure}
\vspace*{13pt}
\centerline{\psfig{figure=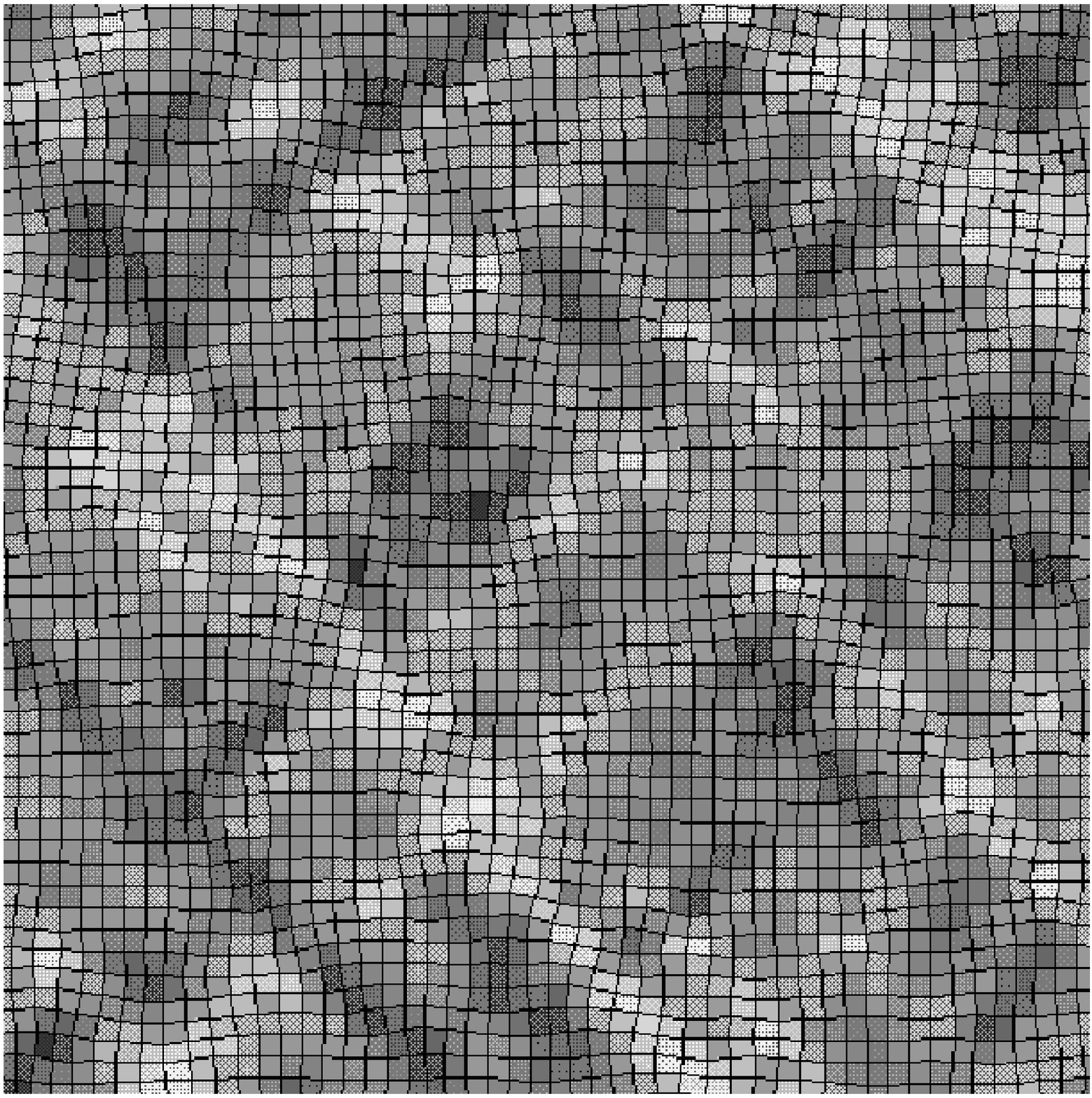,width=4.0truein}}
\bigskip
\fcaption{Tweed as seen in our model.\protect{\cite{Sivan}}
The darkness reflects the amount of diagonal strain
(tall-and-skinny vs.\ short-and-fat). All
materials parameters in our model are determined from independent
experimental measurements in FePd alloys, except for the coupling to
impurities. We set the coupling to impurities to fit the
temperature range for the tweed deformation.}
\label{fig:TheoryTweed}
\end{figure}

We've discussed our model in detail in the literature\cite{PRL,Nobel,Sivan}.
(1)~We blame the tweed on the inescapable, statistical fluctuations in the
local concentration of the atoms in the alloy.  The tweed domains are much
smaller than the martensitic domains in figure~2 (nanometers rather than
microns), and the dependence of the bulk transition temperature on the
average concentration is strong (100K shift every 1\% change in concentration).
(2)~Our numerical model is a Landau--Ginsburg theory, cooled by Monte Carlo:
each little square represents one unit cell.  (3)~The tweed modulations
are due to the elastic anisotropy: the bulk and diagonal shear elastic
constants are much larger than the elastic constant for rectangular
deformations (which mediate the transition).  In the limit of infinite
elastic anisotropy, we show that the only allowed deformations are the
tweedy ones (superpositions of two one-dimensional modulations along the
two diagonals).  (4)~The tweed becomes a thermodynamic phase in this
infinite--anisotropy limit, and is precisely analogous to the frustrated
spin-glass phase in random magnets.

This was an extremely satisfying explanation.  We had claimed all along
that working on spin glasses would one day help explain more practical
materials.  But skepticism in the field remains.  Are the atomic deformations
in tweed large enough for our anharmonic theory to be applicable?  Are they
due to the concentration fluctuations, or perhaps due to local geometries
(large atoms and small atoms straining the local lattice into different
martensitic variants)?  We wanted a clear, dramatic prediction from this
exotic theory.

Glasses are not in equilibrium: their current state depends on the history
of how they are prepared.  Shape-memory alloys have a memory even in the
high-temperature cubic state of strains applied at lower temperatures.
What could be more natural than to store the memory in the glassy tweed state?

Sivan Kartha had gotten his doctorate and moved on to working for clean
cars, so resurrecting the realistic simulation became a major endeavor.
Instead, we set up a simple model in the infinite anisotropy limit (figure~6).
Thus, the fact that the stripes run the full diagonal length of the system is
forced by our model.  We then tried training the tweed: cooling and heating
in a field, then cooling and heating without.  When the strain was
removed, the tweed state returned to a roughly square state --- but
when cooled, the martensite was substantially stretched in the direction
of the original strain.  The tweed had been trained!

\begin{figure}
\vspace*{13pt}
\centerline{\hskip 0.25truein Strained Tweed
                        \hskip 1.25truein Strained Martensite}
\centerline{\hbox to 4.0 truein{
\psfig{figure=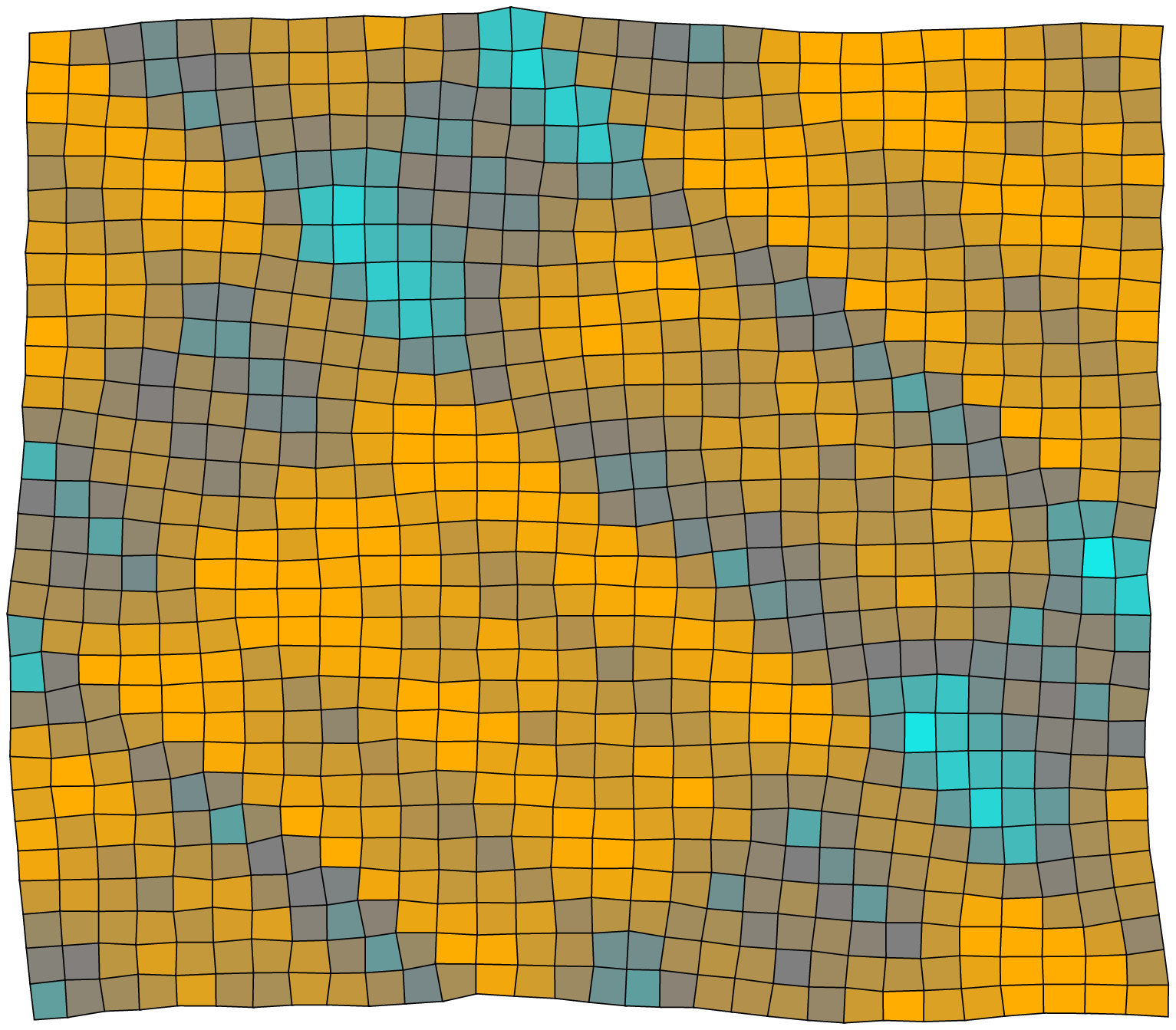,width=1.8truein}\hfil
\psfig{figure=fig/StrainedMartensite.ps,width=1.8truein}}}
\centerline{\hskip 0.25truein Trained Tweed~
                        \hskip 1.25truein Trained Martensite}
\centerline{\hbox to 4.0 truein{
\psfig{figure=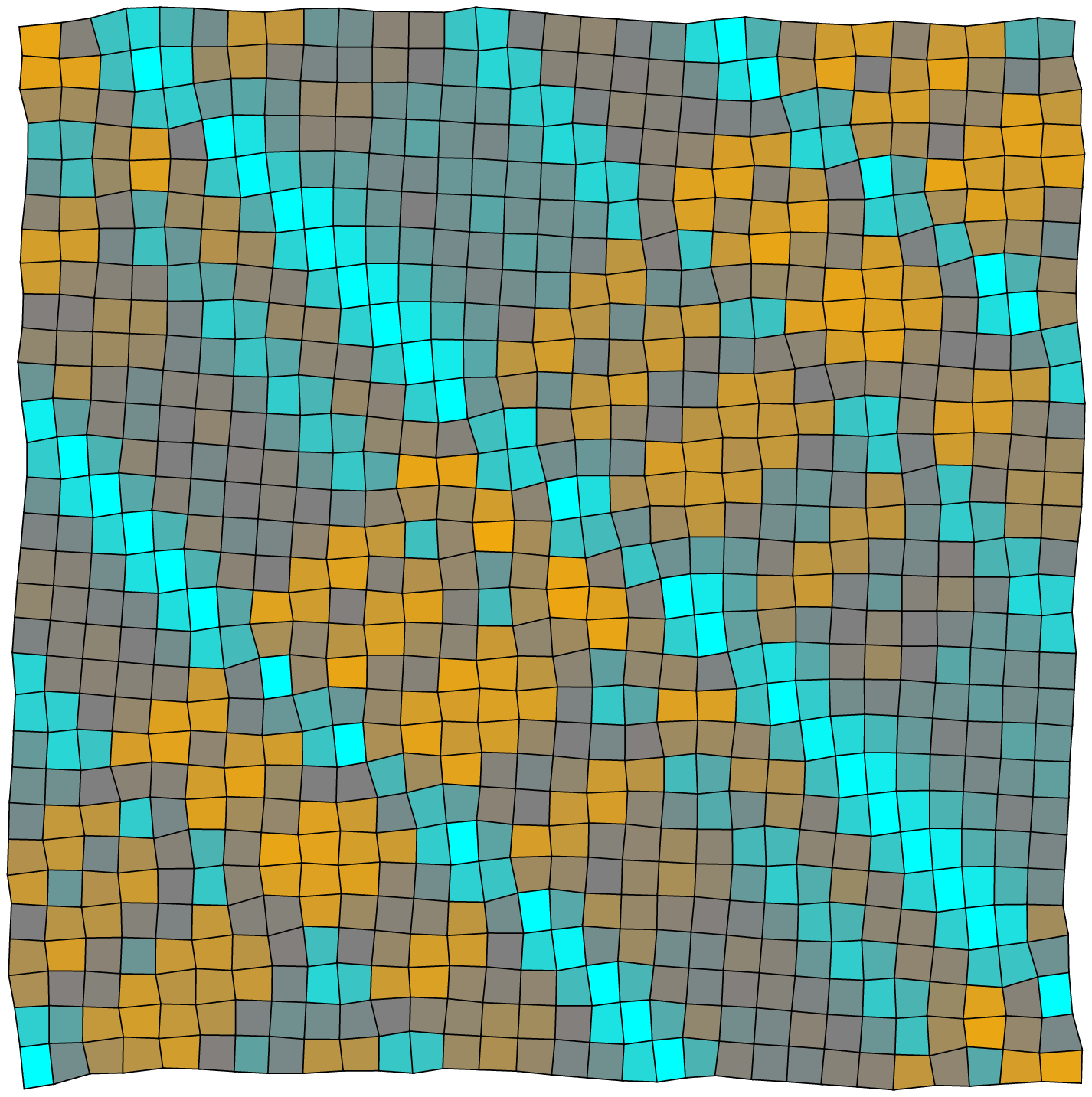,width=1.8truein}\hfil
\psfig{figure=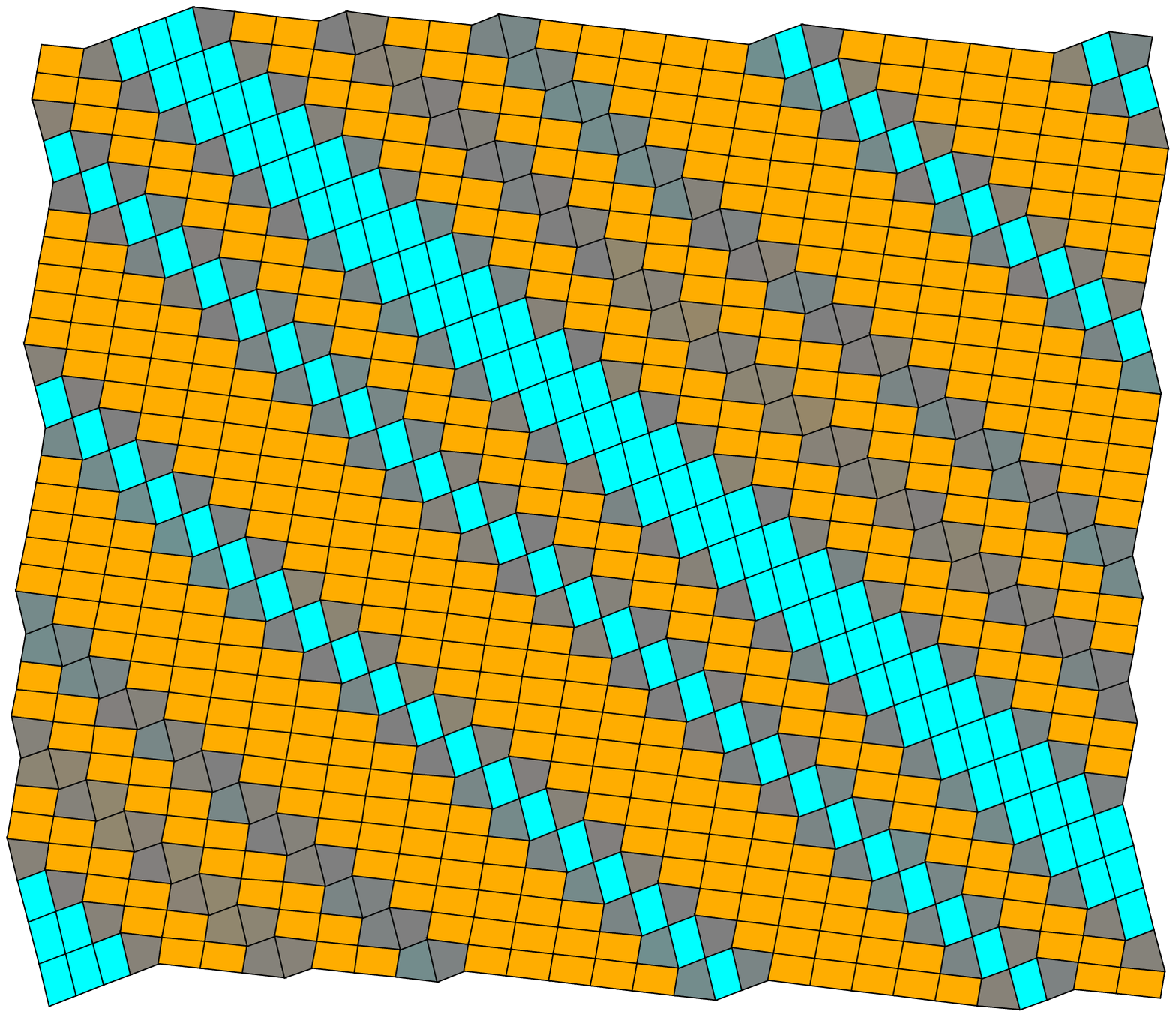,width=1.8truein}}}
\fcaption{{\bf The 2--Way Shape--Memory Effect.} Cycling our
model in an external field strains both the tweed and the martensite
state.  When the field is removed, the tweed springs back into it's
former, roughly cubic shape: the strain seems to have been forgotten.
However, the strain largely reappears when the trained tweed is cooled
into the low-temperature martensitic state.  We propose that this is
the cause of the two-way shape-memory effect.}
\label{fig:Cycle}
\end{figure}

Plotting width versus time for a few runs (figure~7) leads us to believe
that this wasn't an accident.  (1)~The memory improved when we cooled
and heated slowly.  Of course, all deformations are incredibly slow on
atomic time scales. (2)~The transformation occurred in little bursts, or
avalanches, something we've been studying in other contexts.  (3)~The
training was much more rapid than in realistic materials: indeed, the
second cycle often was an exact repetition of the first cycle.  A larger
and more realistic simulation would undoubtedly not repeat exactly after
one cycle, but whether hundreds would be needed is an open question.
Clearly, we should do a better job: we have (slowly) been gearing up to
do simulations of much larger and more realistic systems.

\begin{figure}
\vspace*{13pt}
\centerline{\psfig{figure=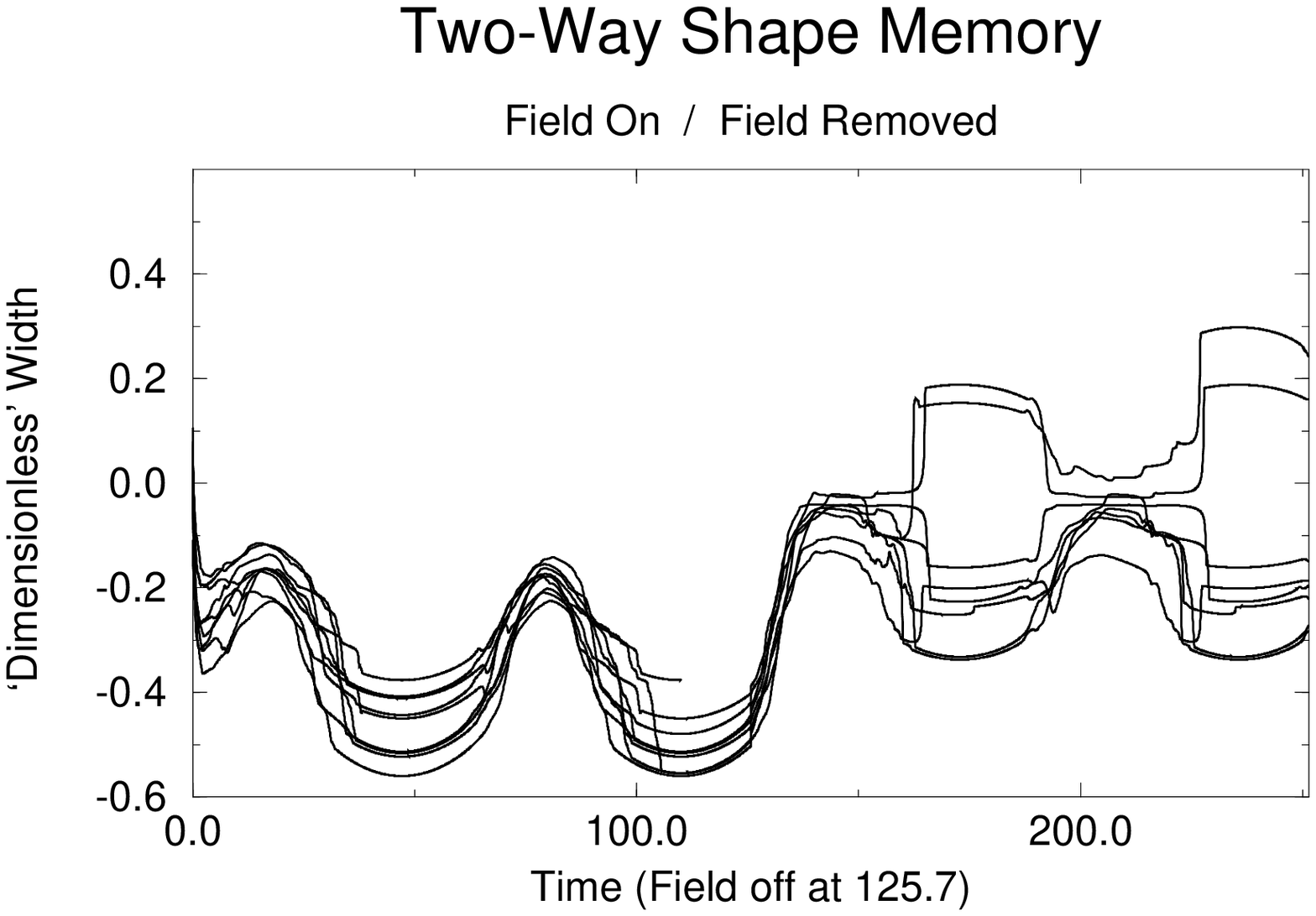,width=4.0truein}}
\bigskip\bigskip\bigskip
\fcaption{{\bf Width versus time.}  The graph shows four
cycles in temperature from high (tweed) to low (martensite); the
first two cycles are in an external field stretching horizontally
(favoring the orange, short-and-fat, negative ``width'' state), the
second two runs are without external field.
Six out of eight runs agree: distort in the direction you've been trained in.}
\label{fig:widthvstime}
\end{figure}

\section{Acknowledgements}
We acknowledge the support of DOE Grant \#DE-FG02-88-ER45364 and
NSF grant \#ASC-9309833, and the Cornell National Supercomputer
Facility, funded in part by NSF, by NY State, and by IBM.
Further pedagogical information using Mosaic is available
at http://www.lassp.cornell.edu/sethna/martensites/WhatIsTweed.html.

\end{document}